\documentclass[sigconf]{arxiv}
\usepackage{times}
\usepackage{soul}
\usepackage{url}
\usepackage{amsmath}
\usepackage{amsthm}
\usepackage{booktabs}
\usepackage{algorithm}
\usepackage{algorithmic}
\usepackage[switch]{lineno}
\usepackage{stfloats}
\usepackage{multirow}
\usepackage{array}
\usepackage{colortbl}
\usepackage{makecell}
\usepackage{threeparttable}
 
\usepackage{amssymb}
\usepackage{xcolor}
\usepackage{subfig}
\usepackage{wrapfig}

\def\ie{{\textit{i}.\textit{e}.}}
\def\etal{{\textit{et al}.}}
\newcommand\Bstrut{\rule[-0.9ex]{0pt}{0pt}} %

\usepackage{marvosym} %

\AtBeginDocument{%
  \providecommand\BibTeX{{%
    \normalfont B\kern-0.5em{\scshape i\kern-0.25em b}\kern-0.8em\TeX}}}

\renewcommand\footnotetextcopyrightpermission[1]{} 

\renewcommand\footnotetextcopyrightpermission[1]{} %
\setcopyright{none}

\acmConference[]{}{}{}

\begin{document}

\title[]{Beyond Score Changes: Adversarial Attack on No-Reference Image Quality Assessment from Two Perspectives}

\author{Chenxi Yang}
\affiliation{%
  \institution{NERCVT, School of Mathematical Sciences, Peking University}
  \city{Beijing}
  \country{China}
  }
\email{yangchenxi@stu.pku.edu.cn}

\author{Yujia Liu}
\affiliation{%
  \institution{NERCVT, School of Computer Science, Peking University}
  \city{Beijing}
  \country{China}
  }
\email{yujia\_liu@pku.edu.cn}

\author{Dingquan Li}
\affiliation{%
  \institution{Peng Cheng Laboratory}
  \city{Shenzhen}
  \country{China}
  }
\email{dingquanli@pku.edu.cn}

\author{Yan Zhong}
\affiliation{%
  \institution{NERCVT, School of Mathematical Sciences, Peking University}
  \city{Beijing}
  \country{China}
  }
\email{zhongyan@stu.pku.edu.cn}

\author[Tingting Jiang]{Tingting Jiang\textsuperscript{\Letter}}
\affiliation{%
  \institution{NERCVT, School of Computer Science, Peking University}
  \city{Beijing}
  \country{China}
  }
\email{ttjiang@pku.edu.cn}

\renewcommand{\shortauthors}{}%
\begin{abstract}
Deep neural networks have demonstrated impressive success in No-Reference Image Quality Assessment (NR-IQA). However, recent researches highlight the vulnerability of NR-IQA models to subtle adversarial perturbations, leading to inconsistencies between model predictions and subjective ratings. Current adversarial attacks, however, focus on perturbing predicted scores of individual images, neglecting the crucial aspect of inter-score correlation relationships within an entire image set. Meanwhile, it is important to note that the correlation, like ranking correlation, plays a significant role in NR-IQA tasks. To comprehensively explore the robustness of NR-IQA models, we introduce a new framework of \textbf{correlation-error-based attacks} that perturb both the correlation within an image set and score changes on individual images. Our research primarily focuses on ranking-related correlation metrics like Spearman's Rank-Order Correlation Coefficient (SROCC) and prediction error-related metrics like Mean Squared Error (MSE). As an instantiation, we propose a practical two-stage \textbf{SROCC-MSE-Attack} (\textbf{SMA}) that initially optimizes target attack scores for the entire image set and then generates adversarial examples guided by these scores.
Experimental results demonstrate that our SMA method not only significantly disrupts the SROCC to negative values but also maintains a considerable change in the scores of individual images. Meanwhile, it exhibits state-of-the-art performance across metrics with different categories. Our method provides a new perspective on the robustness of NR-IQA models.
\end{abstract}

\maketitle
\section{Introduction}
Deep Neural Networks (DNNs) have demonstrated impressive capabilities in diverse domains, as exemplified by their outstanding performance in a range of studies~\cite{2022_CVPRw_MANIQA,2023_CVPR_Chen_AutoFocusFormer,2023_CVPR_Gao_MeMOTR}. Among these, Image Quality Assessment (IQA) stands out as a noteworthy application~\cite{2020_MM_LinearityIQA,2022_CVPRw_MANIQA,ma2023model,zhang2023blind}. The primary objective of IQA is to predict quality scores for input images, aligning with Mean Opinion Scores (MOS) derived from subjective studies, which provide facilities for the evaluation of multimedia tasks~\cite{li2021quality,zhou2022quality}. IQA models can be categorized into Full-Reference (FR) and No-Reference (NR) types based on the availability of reference images: FR-IQA models have access to the reference image, while NR-IQA models do not~\cite{2018_TIP_Bosse_FRNRIQA,2020_TPAMI_Ding_DISTS}. NR-IQA is crucial for designing and optimizing real-world image processing algorithms where the reference image is unavailable, and its growing popularity is attributed to its broader applicability compared to FR-IQA. The evaluation of NR-IQA models is conducted using two major categories of metrics. The first category comprises \textbf{error-based} metrics that measure the prediction error for a single image, such as Mean Squared Error (MSE) and Mean Absolute Error (MAE). The second category is \textbf{correlation-based}, it calculates the correlation relationship between the predicted scores and MOS of a set of images, as observed in metrics like Spearman’s Rank Order Correlation Coefficient (SROCC)~\cite{1961_SROCC}, Kendall Rank Order Correlation Coefficient (KROCC)~\cite{KROCC}, and Pearson Linear Correlation Coefficient (PLCC)~\cite{PLCC}. 

Despite these advances, recent research has unveiled a critical issue – the vulnerability of NR-IQA models to adversarial attacks~\cite{2022_NIPS_Zhang_PAttack,meftah2023evaluating,yang2024exploring,antsiferova2024comparing}. These attacks added subtle perturbations to input images, thus leading to errors in predicted scores while maintaining the appearance of the perturbed image indistinguishable from the original.
We refer to such attacks as \textbf{error-based attacks}.
For instance, Zhang~\etal~\shortcite{2023_NIPS_Zhang_BilossIQA} perturbed low/high-quality images to yield high/low predicted scores, which resulted in a significant Root Mean Squared Error (RMSE) between predicted scores before and after the attacks. Besides, some research has concentrated on the score changes across a set of images. Shumitskaya~\etal~\shortcite{2022_BMVC_Ekarerina_UAP} devised a universal perturbation trained for a given set of images to increase their predicted scores.

\begin{figure}[!t]
    \centering
    \includegraphics[width=\linewidth]{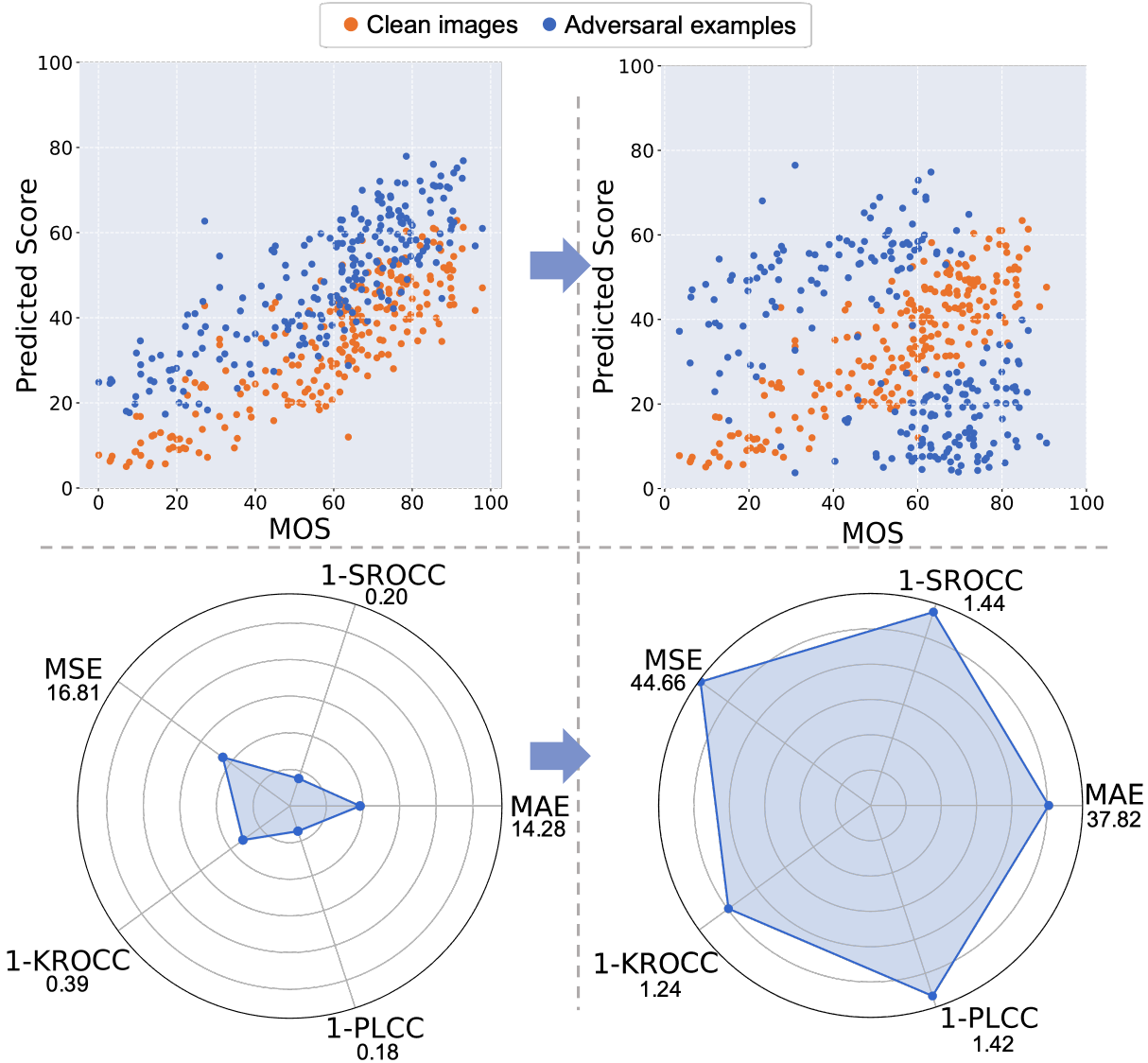}
    \caption{Comparison between error-based attacks (left) and the proposed correlation-error-based attack (right). The first row illustrates the distribution of original scores for clean images and the predicted scores for adversarial examples. Radar charts in the second row assess the attack's performance across five metrics, with each metric calculated between MOS and predicted scores after the attack. A larger polygon area in the radar charts indicates a better overall performance of the attack method.}
    \label{fig:motivation}
\end{figure}

While error-based attacks effectively underscore the vulnerability of NR-IQA models in the variation of the predicted score, these attacks overlook the impact on correlation-based metrics.
As Fig.~\ref{fig:motivation}~(left) shows, error-based attacks may randomly increase 10 to 20 points (total 100) to the predicted score of each sample, resulting in a substantial change in predicted scores, but do not significantly alter the relative ranking of these scores. Consequently, such attacks might be ineffective against correlation-based metrics like SROCC.
Merely targeting the error-based metrics of NR-IQA models is insufficient for a comprehensive attack strategy. An effective attack should take into account all evaluation metrics of the model, including both error-based metrics and correlation-based metrics.
Furthermore, it is important to note that the NR-IQA task is a regression problem where the correlation, like ranking correlation, plays a significant role in its applications. This is particularly evident in applications such as image retrieval~\cite{2020_Signal_Biju_ImageRetri,li2022rankretrieval}. 
Perturbations that affect ranking could seriously compromise retrieval accuracy, underscoring the need to analyze model robustness from the viewpoint of ranking correlation.
In light of these insights, considering attacks that disrupt both error-based and correlation-based metrics becomes a pressing need.

To the best of our knowledge, no existing attack frameworks have specifically addressed the disruption of correlation-based metrics, let alone simultaneously tackled both error-based and correlation-based metrics.
In this paper, to fill this gap, we propose the framework of \textbf{correlation-error-based attacks} for the first time, which demonstrates that attacks of NR-IQA models can effectively impact both error-based and correlation-based metrics concurrently. In this framework, attacks are strategically designed to perturb correlation-based metrics across the entire image set while significantly perturbing the scores for individual samples, as illustrated in Fig.~\ref{fig:motivation}~(right). To achieve these goals, we formulate it as a multi-objective optimization problem from both correlation and error perspectives. For the attack objective, we select MSE and SROCC as the two primary objectives in our multi-objective optimization problem. This selection is based on our observation that different error-based metrics exhibit strong positive correlations, which is similar to correlation-based metrics. Therefore, optimizing MSE and SROCC is expected to bring improvements in other related metrics.

After the problem formulation, We consider how to solve this problem. Directly optimizing MSE and SROCC on the whole image set is challenging due to its high dimensionality. Instead of it, we propose a two-stage \textbf{SROCC-MSE Attack} (\textbf{SMA}) to generate adversarial examples. 
In Stage One, our objective is to identify optimal target scores that significantly reduce SROCC and increase MSE between the predicted scores of attacked and clean images. As SROCC is non-differentiable due to the discrete rank function, we approximate it using a differential function and employ the Lagrangian multiplier method to solve the optimization problem. In Stage Two, we utilize the fast gradient sign method~\cite{2015_ICLR_Goodfellow_FGSM} to generate adversarial examples, striving to make their predicted scores as close as possible to the target scores identified in Stage One. While exact equivalence between the predicted and target scores may not be achievable, our experiments provide compelling evidence that the discrepancy between them is small.

We conduct experiments on the CLIVE dataset~\cite{2015_TIP_LIVEC}, which includes authentic distorted images sourced from the wild and is challenging for NR-IQA. Attacks are conducted on four widely-used NR-IQA models, DBCNN~\cite{2020_TCSVT_DBCNN}, HyperIQA~\cite{2020_CVPR_hyperIQA}, MANIQA~\cite{2022_CVPRw_MANIQA}, and LIQE~\cite{zhang2023blind}. Our approach is compared with four existing attack methods, P-attack~\cite{2022_NIPS_Zhang_PAttack}, OUAP~\cite{2024_CVIU_Shumitskaya_OUAP}, K-attack~\cite{2022_QEVMAw_Korhonen_BIQA}, and UAP~\cite{2022_BMVC_Ekarerina_UAP}.
Our evaluations spanned seven metrics, considering both error and correlation perspectives. The results demonstrate that SMA offers a powerful attack that influences not only the prediction score of individual images but also the correlation within the set of predicted scores. 
For instance, in all attacked models, SMA significantly disrupts the SROCC to negative values while simultaneously maintaining a substantial change in the scores of individual images.
It demonstrates the effectiveness of SMA within our correlation-error-based attacks framework. The findings underscore the vulnerability of NR-IQA in maintaining individual scores and correlations, paving the way for further research on developing more secure and robust NR-IQA models.

To conclude, the contributions of this paper are:
\begin{itemize}
    \item This is the first work to propose the correlation-error-based attack framework, which focuses on both score changes in individual images and the correlation within an image set, filling a gap in the field of attacking correlation-based metrics.
    \item We propose a two-stage SMA method. The Lagrangian multiplier is leveraged to optimize target scores by maximizing score changes (measured by MSE) and disturbing correlation relationships (measured by SROCC). Then individual adversarial images are generated towards target scores. %
    \item  Extensive experiments on four NR-IQA models show the superior performance of SMA in attacking both correlation-based metrics and error-based metrics. It offers a novel perspective on attacks on NR-IQA models, contributing to a more comprehensive attack on NR-IQA models.
\end{itemize}
\section{Related Work}
In this section, we will introduce existing attack methods on NR-IQA models in Sec.~\ref{sec_sample_based}. As a necessary theoretical basis for optimizing ranking property, the differential ranking methods are introduced in Sec.~\ref{sec_diff_rank}.

\subsection{Attacks on NR-IQA Models}\label{sec_sample_based}

Existing attacks on NR-IQA models are all error-based attacks, where attackers aim to perturb error-based metrics like Root Mean Square Error (RMSE). Specifically, error-based attacks involve introducing imperceptible perturbations to a clean image to induce a large change in the score predicted by the NR-IQA model. The resulting perturbed image is termed an adversarial example, and the approach of manipulating input images is commonly referred to as an adversarial attack.

Perceptual attack~\cite{2022_NIPS_Zhang_PAttack} is proposed to attack available NR-IQA models. It changes predicted scores to a certain extent while maintaining the similarity between clean images and adversarial examples, incorporating human feedback in the loop. Korhonen and You~\shortcite{2022_QEVMAw_Korhonen_BIQA} developed a neural network capable of generating adversarial examples, which can be used to attack unknown models. In contrast to the perceptual attack, the universal adversarial perturbation (UAP)~\cite{2022_BMVC_Ekarerina_UAP} is designed to execute attacks by considering a set of images. UAP aimed at generating a universal perturbation applicable to a set of images, to collectively increase the predicted scores by an NR-IQA model. It implies the adversarial perturbation remains the same for all input images within the set. Similar to UAP, OUAP~\shortcite{2024_CVIU_Shumitskaya_OUAP} also generates a universal perturbation for a set of images, maximizing the score predicted by the attacked NR-IQA model on a set of images. However, the imperceptibility of universal perturbations is not satisfactory.

It is important to note that existing methods focus on modifying the error of predicted scores without taking into account the correlation relationship between predicted scores of adversarial examples and original scores. This paper studies this aspect and offers a more comprehensive perspective to investigate the robustness of NR-IQA models.

\subsection{Differential Ranking}\label{sec_diff_rank}
Consider a set of image quality scores denoted as $\boldsymbol{s} = (s_1, s_2, \dots, s_n)$. The ranking function, defined as
\begin{equation}
r: \boldsymbol{s} \rightarrow \text{Perm}(n),
\end{equation}
maps the set of scores to a permutation of $\{1,2,\ldots, n\}$. It evaluates at the index $j$ to determine the position of $s_j$ in the descending sort (a smaller rank $r_j(\boldsymbol{s})$ indicates a higher value for $s_j$). To illustrate, if $s_1 > s_3 > s_2$, then $r(\boldsymbol{s}) = (1, 3, 2)$. Notably, the function $r$ is piecewise constant, rendering its derivatives either null or undefined. Consequently, this characteristic hinders the application of gradient backpropagation when we want to optimize the ranking metrics.

Many studies have explored alternative approaches to handle differentiable proxies for ranking. One feasible way is directly approximating ranking metrics, such as minimizing a smooth approximation of ranking measures through gradient descent strategies~\cite{2010_Chapelle_DirectRank} or employing diverse forms of loss-augmented inference~\cite{2018_CVPR_Mohapatra_EfficientRank}. Another prevalent method is referred to as ``soft'' ranking, which can be integrated into any differential loss function. Various techniques have been employed to achieve this goal. For instance, Taylor~\etal~\shortcite{2008_ICWSDM_taylor_softrank} introduced a method where each score is smoothed using equal variance Gaussian distributions. Engilberge~\etal~\shortcite{2019_CVPR_Engilberge_SoDeep} took a different approach by training a deep neural network to learn the entire sorting operation. However, these methods tend to be time-consuming. In contrast, Blondel~\etal~\shortcite{2020_ICML_Blondel_FastRank} proposed an efficient approach by constructing differentiable operators as projections onto the convex hull of permutations, demonstrating speed and computational friendliness.
In this paper, we adopt the method proposed by Blondel~\etal~\shortcite{2020_ICML_Blondel_FastRank} during the SROCC optimization phase.
\section{Methodology}
In this section, Sec.~\ref{sec:prob_def} introduces the definition of correlation-error-based attacks on NR-IQA models, with a particular focus on the problem of multiple objective optimization across both correlation-based and error-based metrics. Based on this optimization problem, we propose a two-stage SROCC-MSE Attack (SMA) method, and the overall idea is illustrated in Sec.~\ref{sec:method_overall}. In Sec.~\ref{sec:stage_1}, we demonstrate how to get optimal target scores in Stage One. Finally, Sec.~\ref{sec:stage_2} explains how these target scores will guide the generation of adversarial examples in the targeted samples generation stage.

\subsection{Problem Definition}\label{sec:prob_def}
In this paper, we consider correlation-error-based attacks on an NR-IQA model. Attackers design attack strategies based on a set of images and the goal is to perturb both the correlation relationship and score changes between predicted scores of adversarial examples and original scores of clean images. For instance, attackers may want to lead large score changes on individual images, meanwhile resulting in an inconsistency between the rankings of predicted scores and original scores.

Consider an image set $\mathcal{I} = \{I_j\}_{j=1}^N$ containing $N$ images and $f(\cdot)$ is an NR-IQA model. The original score of image $I_j$ predicted by $f(\cdot)$ is $s_j=f(I_j)$, and we use $\boldsymbol{s}:=f(\mathcal{I})=\{s_j\}_{j=1}^N$ to represent the set of original scores. Let $I'_j$ be the perturbed version of $I_j$ and the set of adversarial examples is denoted as $\mathcal{I}' = \{I'_j\}_{j=1}^N$. The set of predicted scores after attacks is represented as $\boldsymbol{s}':=f(\mathcal{I}')=\{s'_j\}_{j=1}^N$. Therefore, the optimization problem for correlation-error-based attacks is formulated as
\begin{equation}
\begin{split}
    \min_{\mathcal{I}'} ~~ \mathcal{L}_\text{cor}& (f(\mathcal{I}'), f(\mathcal{I}))-\lambda\mathcal{L}_\text{err} (f(\mathcal{I}'), f(\mathcal{I})), \\
    \text{s.t.  } & D(I'_j, I_j) \leqslant \epsilon, \forall j \in \{1,\cdots,N\}.
\end{split}
\end{equation}
The correlation loss function, denoted as $\mathcal{L}_\text{cor}(\cdot, \cdot)$, is formulated based on two sets of scores, aiming to evaluate the correlation between them.
The error loss function, $\mathcal{L}_\text{err}(\cdot, \cdot)$, is constructed for each pair of scores $\{f({I}_j'),f({I}_j)\}_{j=1}^N$, quantifying the prediction error between $f({I}_j')$ and $f({I}_j)$. %
The positive $\lambda$ measures the trade-off between two optimization objectives of correction and error.
And $D(\cdot, \cdot)$ calculates the perceptual distance between $I'_j$ and $I_j$ with a pre-defined threshold $\epsilon$.
We assume that as long as $I'_j$ and $I_j$ satisfy the condition $D(I'_j, I_j) \leqslant \epsilon$, they do not appear to have a significant quality difference to human eyes so that they should have the same subjective quality perceived by humans.

In this paper, we focus on the $\mathcal{L}_\text{cor}(\cdot, \cdot)$ related to the ranking, and mainly consider the SROCC~\cite{1961_SROCC} commonly used in IQA tasks~\cite{siniukov2022applicability,madhusudana2022image,siniukov2023unveiling}, \ie,
\begin{equation}
\begin{aligned}
    \mathcal{L}_\text{cor} (\boldsymbol{s}', \boldsymbol{s}) &= 
    \text{SROCC}(\boldsymbol{s}', \boldsymbol{s}) \\ &=
    1 - \frac{6 }
    {N(N^2 - 1)}\sum\limits_{j=1}^N \left(r_j(\boldsymbol{s}') - r_j(\boldsymbol{s})\right)^2.
\end{aligned}
\end{equation}
Here, $r_j(\cdot)$ is the $j^{th}$ element of the ranking function $r(\cdot)$. For $\mathcal{L}_\text{err}(\cdot, \cdot)$, we leverage MSE to measure the error between $s'_j$ and $s_j$:
\begin{equation}
    \mathcal{L}_\text{err} (\boldsymbol{s}', \boldsymbol{s}) = \text{MSE}(\boldsymbol{s}', \boldsymbol{s}) = \frac{1}{N}\sum_{j=1}^N\left(s'_j- s_j\right)^2.
\end{equation}

The perceptual distance $D(\cdot, \cdot)$ has many choices~\cite{2022_NIPS_Zhang_PAttack}, such as the $\ell_\infty$ norm, SSIM~\cite{2004_TIP_Wang_SSIM}, LPIPS~\cite{2018_CVPR_Richard_LPIPS}, and so on~\cite{sheikh2006image,reisenhofer2018haar,2020_TPAMI_Ding_DISTS}.
Among them, we choose the most commonly used $\ell_\infty$ norm as $D(\cdot, \cdot)$ for its simplicity in computation and optimization.

To conclude, in this paper, we aim to generate adversarial examples by solving the following optimization problem:
\begin{equation}
\begin{split}
\label{eq:objective_function}
    \min_{\mathcal{I}'}~~\text{SROCC}&(f(\mathcal{I}'), f(\mathcal{I}))-\lambda_\text{MSE}\text{MSE}(f(\mathcal{I}'), f(\mathcal{I})), \\
    \text{s.t.  } & \left\Vert I'_j - I_j \right\Vert_\infty \leqslant \epsilon, \forall j \in \{1,\cdots,N\},
\end{split}
\end{equation}
where $\lambda_\text{MSE}>0$ is the weight to balance two attack objectives of correlation- and error-based metrics.

\begin{figure*}[!t]
    \centering
    \includegraphics[width=\textwidth]{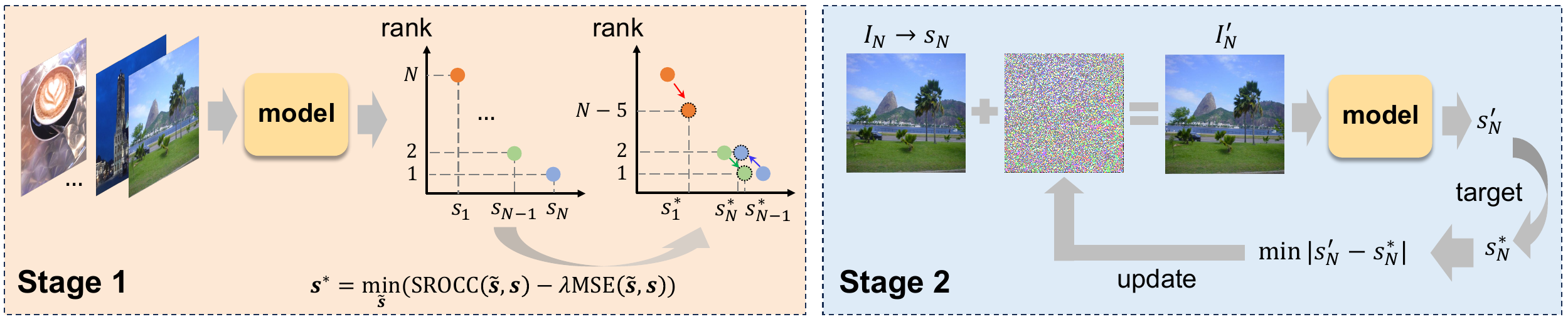}
    \caption{Overall methodology of the proposed SMA method.}
    \label{fig:overall}
\end{figure*}

\subsection{Overall Methodology}\label{sec:method_overall}
A straightforward approach to address the optimization problem presented in Eq.~\eqref{eq:objective_function} is to solve it directly. However, there are two challenges. The first challenge is the storage consumption. Solving Eq.~\eqref{eq:objective_function} directly requires the entire set $\mathcal{I}$ to be input into $f(\cdot)$, thereby demanding gradients for every image in the set during the optimization phase. This requirement to store gradients for all images incurs considerable overhead. The second challenge is the high dimensionality of the problem defined in Eq.~\eqref{eq:objective_function}. Since the optimal solution of Eq.~\eqref{eq:objective_function} is a set of adversarial examples, the dimensionality of this problem is $N\times d$, where $d$ is the dimension of a single image. In many real-world applications, $\mathcal{I}$ comprises an extensive number of large-scale images, making the direct solution of such a complex optimization problem impractical.

To overcome these challenges, we try to decompose this $N\times d$-dimensional problem into $N$ $d$-dimensional optimization problems guided by the information of $f(\mathcal{I})$. To achieve this goal, we propose a two-stage approach as illustrated in Fig.~\ref{fig:overall}. In Stage One, at the score level, we find a set of target scores $\boldsymbol{s}^*=\{s_1^*, \dots, s_N^*\}$ which can minimize the value of $\text{SROCC}(\boldsymbol{s}^*,\boldsymbol{s})-\lambda_\text{MSE}\text{MSE}(\boldsymbol{s}^*,\boldsymbol{s})$. In Stage Two, at the image level,  for each image $I_j\in\mathcal{I}$, the optimization problem is to generate an adversarial example $I'_j$ such that its predicted score is close to $s^*_j$ as much as possible. By this two-stage approach, we decompose the original problem into two levels, score level, and image level, which makes it easier to handle. The next two sections introduce the two stages of the proposed SMA method in detail.

\subsection{Target Scores Optimization}\label{sec:stage_1}
The objective of the target scores optimization stage in SMA is to determine a set of target scores
\begin{equation}
\label{eq:min_srocc}
    \boldsymbol{s}^*=\mathop{\arg\min}_{\boldsymbol{\tilde{s}}} 
    \text{SROCC}(\boldsymbol{\tilde{s}}, \boldsymbol{s})-\lambda_\text{MSE}\text{MSE}(\boldsymbol{\tilde{s}}, \boldsymbol{s}),
\end{equation}
where $\boldsymbol{s}=\{f(I_j)\}_{j=1}^N$ represents the set of original predicted scores for the image set $\mathcal{I}=\{I_j\}_{j=1}^N$, $\lambda_\text{MSE}>0$ provides a trade-off between SROCC and MSE.

We find that there is a problem of $\boldsymbol{s}^*$ in guiding the sample generation in Stage Two if we only minimize SROCC and maximize MSE as Eq.~\eqref{eq:min_srocc} shows. Specifically, since it is hard for Stage Two to generate adversarial examples with precisely predicted scores $\boldsymbol{s}^*$, the difference between the final predicted scores $\boldsymbol{s}'$ and the target scores $\boldsymbol{s}^*$ can potentially disrupt the ranking results $r(\boldsymbol{s}^*)$. This mismatch may result in a significant gap between the actual $\text{SROCC}(\boldsymbol{s}',\boldsymbol{s})$ and the ideal $\text{SROCC}(\boldsymbol{s}^*,\boldsymbol{s})$. For example, given three images whose original scores $\boldsymbol{s}=\{3,2,1\}$, the optimal target scores calculated in Stage One are $\boldsymbol{s}^*=\{4.99,5.00,5.01\}$, and the predicted scores of adversarial examples generated in Stage Two are $\boldsymbol{s}'=\{5.00, 4.99, 4.98\}$. Despite a small bias between $\boldsymbol{s}^*$ and $\boldsymbol{s}'$ (RMSE value is less than $0.03$), the actual $\text{SROCC}(\boldsymbol{s}',\boldsymbol{s})=1$ significantly differs from the ideal $\text{SROCC}(\boldsymbol{s}^*,\boldsymbol{s})=-1$.

To solve this problem, we introduce an interior-constraint strategy for target scores $\boldsymbol{s}^*$. We maximize the variance of $\boldsymbol{s}^*$ so that the distribution of scores $\{s^*_1,\dots,s^*_N\}$ remains as dispersed as possible. This strategy is helpful for Stage Two, as it enhances the tolerance of bias between predicted scores and target scores. Continuing with the previous example, if target scores are $\boldsymbol{s}^*=\{0,3.99,5.01\}$, then as long as the final predicted scores $\boldsymbol{s}'$ satisfy $|s'_j-s^*_j| < 0.51$, $\text{SROCC}(\boldsymbol{s}',\boldsymbol{s})$ is equivalent to $\text{SROCC}(\boldsymbol{s}^*,\boldsymbol{s})$.

Therefore, the optimization problem transforms into finding the $\boldsymbol{s}^*$ that minimizes the expression:
\begin{equation}\label{eq:stage1}
    \boldsymbol{s}^*=\mathop{\arg\min}_{\boldsymbol{\tilde{s}}} 
    \left(
    \text{SROCC}(\boldsymbol{\tilde{s}}, \boldsymbol{s}) - \lambda_\text{Var} \text{Var}(\boldsymbol{\tilde{s}}) - \lambda_\text{MSE}\text{MSE}(\boldsymbol{\tilde{s}}, \boldsymbol{s})
    \right),
\end{equation}
where $\lambda_\text{Var},\lambda_\text{MSE}>0$ represent the weight balances among the SROCC, the tolerance of bias, and the MSE. And 
\begin{equation}
\label{eq:stage1_var}
    \text{Var}(\boldsymbol{\tilde{s}}) = \frac{1}{N-1} \sum_{j=1}^N \left(
    \tilde{s}_j - 
    \frac{1}{N}\sum_{k=1}^N \tilde{s}_k
    \right)^2
\end{equation}
is the variance of $\boldsymbol{\tilde{s}}$.

\begin{figure}[!t]
    \centering \includegraphics[width=0.72\linewidth]{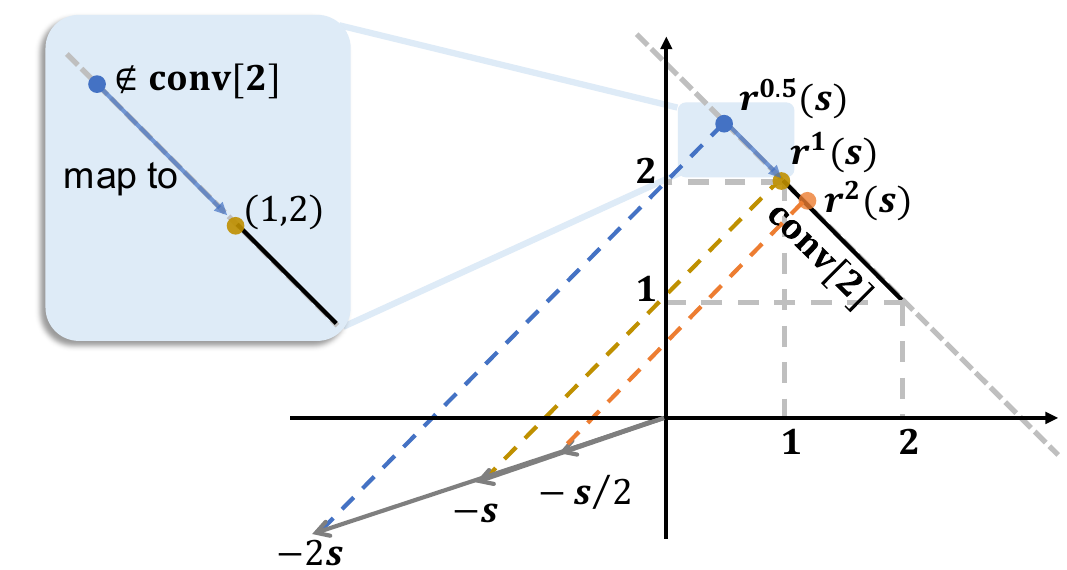}
    \caption{An example of the approximation function $r^\beta(\boldsymbol{s})$ which calculates the Euclidean projection of $\boldsymbol{s}/\beta$ on to $\textbf{conv}[N]$. In this example, $\boldsymbol{s}=\{1.5, 0.5\}$ is represented by a vector $(1.5,0.5)\in\mathbb{R}^2$. }
    \label{fig:soft_rank}
\end{figure}

The challenge in solving Eq.~\eqref{eq:stage1} arises from the non-differentiable ranking function $r(\cdot)$ when calculating $\text{SROCC}(\cdot,\cdot)$. Recall that
$ \text{SROCC}(\boldsymbol{\tilde{s}}, \boldsymbol{s}) =
 1 - \frac{6}    {N(N^2 - 1)}\sum_{j=1}^N \left(r_j(\boldsymbol{\tilde{s}}) - r_j(\boldsymbol{s})\right)^2 $.
It poses a challenge for optimizing the objective function with gradient-based algorithms due to the non-differentiability.

To solve the optimization problem defined in Eq.~\eqref{eq:stage1} with gradient-based methods, we approximate the non-differential ranking function $r(\cdot)$ with a differential function $r^\beta(\cdot)$ where $\beta > 0$ serves as a hyper-parameter. Following the approach of Blondel \etal~\shortcite{2020_ICML_Blondel_FastRank}, $r(\cdot)$ is approximated by
\begin{equation}
\label{eq:rank_approxi}
    r^\beta(\boldsymbol{s}) = \mathop{\arg\min}_{\boldsymbol{z} \in \textbf{conv}[N]}
    \frac{1}{2} \left\Vert \boldsymbol{z} + \frac{\boldsymbol{s}}{\beta} \right\Vert^2,
\end{equation}
where $\textbf{conv}[N]$ is a convex polytope whose vertices correspond to permutations in $[N]$. The notation $[N]$ represents all permutations of $\{1,\dots,N\}$ which contain $N!$ vertices\footnote{[N]=\{(1,2,3\dots,N),(2,1,3,\dots,N),\dots,(N,N-1,N-2\dots,1)\}.}. As illustrated in Fig.~\ref{fig:soft_rank}, $r^\beta(\boldsymbol{s})$ calculates the Euclidean projection of $\boldsymbol{s}/\beta$ on to $\textbf{conv}[N]$. If the projection of $\boldsymbol{s}/\beta$ lies outside $\textbf{conv}[N]$, $r^\beta(\boldsymbol{s})$ computes the nearest vertex of $\textbf{conv}[N]$ to the projection. The function $r^\beta(\cdot)$ is differential, as proven in the work~\cite{2020_ICML_Blondel_FastRank}.
Furthermore, the approximation error, measured by $\Vert r^\beta(\boldsymbol{s}) - r(\boldsymbol{s})\Vert_2^2$, is close to zero when $\beta$ is sufficiently small~\cite{2020_ICML_Blondel_FastRank}. As a special case in \shortcite{2020_ICML_Blondel_FastRank}, we have:
\begin{theorem}
\label{thr:approximation_error}
For all $\boldsymbol{s} \in \mathbb{R}^N$ without ties, $r^\beta(\boldsymbol{s}) = r(\boldsymbol{s})$ when 
\begin{equation}
    \beta \leqslant \min_{i\in\{1,\dots,N-1\}} \boldsymbol{s}_{r^{-1}(i)} - \boldsymbol{s}_{r^{-1}(i+1)}.
\end{equation}
Here, $r^{-1}(\cdot)$ represents the inverse function of $r(\cdot)$, where $r^{-1}(i)$ denotes the index of the $i^{th}$ largest element in $\boldsymbol{s}$.
\end{theorem}

Therefore, we replace the ranking function in SROCC with $r^\beta(\cdot)$ to make the objective function differentiable. 
The modified objective function is expressed as:
\begin{equation}
\label{eq:objective_stage_1}
\begin{aligned}
    \min_{\tilde{s}}
    1 - \frac{6} {N(N^2 - 1)}\sum_{j=1}^N \left(r^\beta_j(\boldsymbol{\tilde{s}}) - r^\beta_j(\boldsymbol{s})\right)^2 \\
    - \lambda_\text{Var}\text{Var}(\boldsymbol{\tilde{s}})-\lambda_\text{MSE}\text{MSE}(\boldsymbol{\tilde{s}},\boldsymbol{s}).
\end{aligned}
\end{equation}

\subsection{Targeted Samples Generation}\label{sec:stage_2}
In Stage Two of SMA, for a given original image $I_j (j=1,\dots,N)$, the objective is to generate an adversarial example $I'_j$ with a predicted score $s'_j$ as close to the target score $s^*_j$ as possible. Meanwhile, $I'_j$ is required to satisfy $\Vert I'_j - I_j \Vert_\infty \leqslant \epsilon$ where $\epsilon$ is a pre-defined threshold, ensuring that the adversarial perturbation remains imperceptible to human eyes. Therefore, the optimization problem for each input image is expressed as follows:
\begin{equation}
\label{eq:second_stage}
    \min_{I'_j} \left( f(I'_j) - s^*_j\right)^2,~~~~\text{s.t. } \left\Vert I'_j - I_j \right\Vert_\infty \leqslant \epsilon.
\end{equation}
In this equation, $f(\cdot)$ represents the NR-IQA model to be attacked.

To efficiently generate $I'_j$ satisfying Eq.~\eqref{eq:second_stage}, we draw inspiration from a classical attack method in the classification task --- the fast gradient sign method (FGSM)~\cite{2015_ICLR_Goodfellow_FGSM}.
We adopt an iterative approach to update the adversarial sample, incorporating the clamp function to ensure that $\Vert I'_j - I_j \Vert\infty \leqslant \epsilon$.

In detail, we initialize the adversarial example $I'_j$ as $I_j$ and update $I'_j$ iteratively using the following formula:
\begin{equation}
    I'_j \leftarrow I'_j - \alpha \cdot \text{sign} \left( 
    \nabla_{I'_j} \left( f(I'_j) - s^*_j\right)^2
    \right),
\end{equation}
where $\alpha$ is the step size, and $\text{sign}(\cdot)$ is the sign function. This update operation is repeated for $m$ iterations.
Additionally, after each update of $I'_j$, we clamp its pixel values to the range $[I_j - \epsilon, I_j + \epsilon]$ to satisfy the constraint condition in Eq.~\eqref{eq:second_stage}. The complete algorithm is outlined in Algorithm~\ref{alg:second_stage}.

\begin{algorithm}[tb]
    \caption{Stage Two of SMA}
    \label{alg:second_stage}
    \textbf{Input}: NR-IQA model $f(\cdot)$, original image $I_j$, target score $s^*_j$, pre-defined threshold $\epsilon$ for $\ell_\infty$ norm of the perturbation, iteration number $m$, step size $\alpha$\\
    \textbf{Output}: Adversarial example $I'_j$
    \begin{algorithmic}[1] %
        \STATE Let $k\leftarrow 1$, $I^0_j \leftarrow I_j$
        \WHILE{$k \leqslant m$}
        \STATE $\mathcal{L}_{\text{adv}} \leftarrow \left( f(I^{k-1}_j) - s^*_j\right)^2$
        \STATE $I^k_j \leftarrow I^{k-1}_j + \alpha \cdot \text{sign} \left( \nabla_{I^{k-1}_j} \mathcal{L}_{\text{adv}} \right)$
        \IF {$\Vert I^k_j - I_j \Vert_\infty > \epsilon$}
        \STATE Clamp pixels in $I^k_j$ within the range $[I_j - \epsilon, I_j + \epsilon]$.
        \ENDIF
        \STATE $k \leftarrow k + 1$
        \ENDWHILE
        \STATE \textbf{return} $I'_j \leftarrow I^m_j$
    \end{algorithmic}
\end{algorithm}
\section{Experiments}
In this section, we present the experimental results conducted on the CLIVE~\cite{2015_TIP_LIVEC} dataset, which collected images distorted in the wild and is challenging for the NR-IQA task. We begin by comparing the proposed SMA method with four state-of-the-art (SOTA) attack methods on four NR-IQA models in Sec.~\ref{sec:exper_sota}. Then we analyze the insight provided by the optimization strategy used in Stage One in Sec.~\ref{sec:exper_stage_1}. We discuss the error between the target scores and predicted scores in Stage Two in Sec.~\ref{sec:exper_error}. %
Ablation studies that explore hyperparameters in this work are presented in Sec.~\ref{sec:exper_ablation}.

\subsection{Comparison with SOTA Attacks}
\label{sec:exper_sota}
We compare our method with four NR-IQA attack methods: the perceptual attack (P-attack)~\cite{2022_NIPS_Zhang_PAttack}, OUAP~\cite{2024_CVIU_Shumitskaya_OUAP}, the attack method proposed by Korhonen and You (K-attack)~\shortcite{2022_QEVMAw_Korhonen_BIQA}, and UAP~\cite{2022_BMVC_Ekarerina_UAP}. Among them, K-attack and UAP generate adversarial examples on a substitute model and then transfer these samples to the target NR-IQA model. We select four widely-used NR-IQA models as the target model: DBCNN~\cite{2020_TCSVT_DBCNN}, HyperIQA~\cite{2020_CVPR_hyperIQA}, MANIQA~\cite{2022_CVPRw_MANIQA}, and LIQE~\cite{zhang2023blind}, whose architectures are based on ResNet~\cite{2016_CVPR_He_ResNet}, VGG~\cite{vgg}, transformer ViT~\cite{dosovitskiy2020vit} and CLIP~\cite{alec2021learning}, respectively. For all models, $80\%$ of the images are randomly selected from the CLIVE dataset for training NR-IQA models, with the remaining images (a total of $232$ images) for testing. During the attack phase, only the test images are used. According to the original settings of these models, images with full size $500\times 500$ are the inputs of DBCNN, and images are center-cropped to the size of $224\times 224$ before being input to other three NR-IQA models.
Additionally, we add a clipping function into the final layer of all NR-IQA models to ensure that predicted scores fall within the range of $[0, 100]$. %

\begin{figure}[!t]
    \centering
    \includegraphics[width=0.9\linewidth]{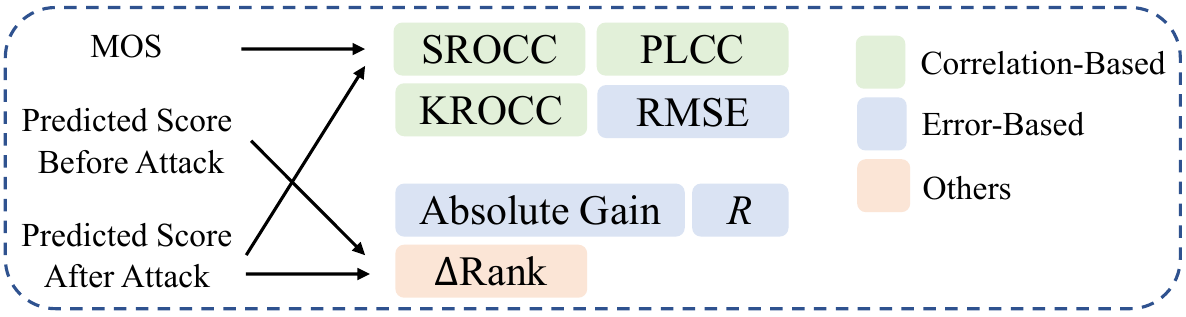}
    \caption{Evaluation metrics utilized in our experiments. Correlation-Based/ Error-Based/Other metrics are marked with different colors.}
    \label{fig:metrics}
    \vspace{-2ex}
\end{figure}

To comprehensively evaluate the attack performance from different perspectives, seven metrics are considered, as illustrated in Fig.~\ref{fig:metrics}. SROCC, KROCC, PLCC, and RMSE are conducted between predicted scores of adversarial examples and MOS provided by humans. In general, smaller values of SROCC, KROCC, and PLCC, as well as a larger value of RMSE, indicate a stronger attack. The Absolute Gain~\cite{antsiferova2024comparing} calculated the MAE between the predicted score before and after the attack, which is defined as $\text{Abs.} = \frac{1}{N}\sum_{j=1}^N|f(I_j')-f(I_j)|$,
where $f(\cdot)$ is the attacked NR-IQA model, $I_j$ and $I_j'$ are $j^\text{th}$ image and its adversarial example respectively. 
The $R$~\cite{2022_NIPS_Zhang_PAttack} is defined as:
\begin{equation}
    R = \frac{1}{N}\sum_{j=1}^N \log\left(\frac{\max\{\beta_1-f(I_j),f(I_j)-\beta_2\}}{|f(I_j)-f(I_j')|}\right),
\end{equation}
where $\beta_1$ and $\beta_2$ are the maximum MOS and minimum MOS among all MOS values. The $\beta_1,\beta_2$ are set to $3.50, 90.55$ in our experiments.
The $\Delta$Rank measures the change in the rank of each image before and after the attack, defined as:
\begin{equation}
    \Delta\text{Rank} = \frac{1}{N}\sum_{j=1}^{N}\left|r_j(f(\mathcal{I'}))-r_j(f(\mathcal{I}))\right|,
\end{equation}
where $r_j(f(\mathcal{I}))$ and $r_j(f(\mathcal{I'}))$ are the rank of the predicted score of $j^\text{th}$ image before/after the attack within the image set. $\Delta\text{Rank}$ provides a vital reference for the recommendation systems where the ranking of images is important. 
Generally, larger values of Abs., $\Delta$Rank, and a smaller value of $R$ correspond to a stronger attack.

Meanwhile, we consider the imperceptibility of adversarial perturbations, ensuring that the MOS of adversarial examples is the same as that of clean images. 
We employ the commonly-used metrics, LPIPS~\cite{2018_CVPR_Richard_LPIPS}, to quantify the visual similarity between adversarial examples and clean images. Smaller LPIPS values indicate more imperceptible perturbations. 

In our experiments, we use $r^1(\cdot)$ to approximate the ranking function. $\lambda_\text{Var}, \lambda_\text{MSE}$ in Eq.~\eqref{eq:objective_stage_1} are chosen as $0.0001$  for DBCNN, HyperIQA and LIQE. For MANIQA, $\lambda_\text{Var}, \lambda_\text{MSE}$ are set to $0.0002$ and $0.00001$, respectively. The objective in Eq.~\eqref{eq:objective_stage_1} is optimized $100,000$ epochs with Adam optimizer~\cite{kingma2015adam}.
The $\epsilon$ in Eq.~\eqref{eq:second_stage} is set to $0.005$. The iteration number $m$ and step size $\alpha$ are set to $10, 0.005$, respectively. The comparative results are shown in Table~\ref{tab:sota_compare}.

\begin{table*}[!t]
\setlength\tabcolsep{1.2pt} %
\centering
\caption{Comparison with SOTA methods on attacking NR-IQA models DBCNN, HyperIQA, MANIQA and LIQE (\textbf{Bold}/ \underline{underlined} denotes the best/ the second best value in each column)}
\label{tab:sota_compare}
\resizebox{\textwidth}{!}{
\begin{tabular}{
l>{\centering\arraybackslash}p{1.1cm}>{\centering\arraybackslash}p{1.1cm}>{\centering\arraybackslash}p{1.1cm}>{\centering\arraybackslash}p{1.1cm}|>{\centering\arraybackslash}p{1.1cm}>{\centering\arraybackslash}p{1.1cm}>{\centering\arraybackslash}p{1.1cm}|>{\centering\arraybackslash}p{1.1cm}|>{\centering\arraybackslash}p{1.1cm}>{\centering\arraybackslash}p{1.1cm}>{\centering\arraybackslash}p{1.1cm}>{\centering\arraybackslash}p{1.1cm}|>{\centering\arraybackslash}p{1.1cm}>{\centering\arraybackslash}p{1.1cm}>{\centering\arraybackslash}p{1.0cm}|>{\centering\arraybackslash}p{1.1cm}}
\toprule
& \multicolumn{8}{c|}{DBCNN}               & \multicolumn{8}{c}{HyperIQA}          \\ \cmidrule(lr){2-17}
 & \multicolumn{4}{c|}{Predicted vs. MOS}                  & \multicolumn{3}{c|}{Before vs. After Attack} & \multicolumn{1}{c|}{Invis.}       & \multicolumn{4}{c|}{Predicted vs. MOS} & \multicolumn{3}{c|}{Before vs. After Attack} & \multicolumn{1}{c}{Invis.}   \\
& SROCC$\downarrow$                    & KROCC$\downarrow$                    & PLCC$\downarrow$                     & \multicolumn{1}{c|}{RMSE$\uparrow$}                      & Abs.$\uparrow$    & $R\downarrow$ & $\Delta$Rank$\uparrow$        & LPIPS$\downarrow$                  & SROCC$\downarrow$                   & KROCC$\downarrow$                   & PLCC$\downarrow$                    & \multicolumn{1}{c|}{RMSE$\uparrow$} & Abs.$\uparrow$    & $R\downarrow$      & $\Delta$Rank$\uparrow$ & LPIPS$\downarrow$                  \\ \hline
Clean & 0.868 & 0.889 & 0.686 & 10.782 & - & - & - & -            & 0.879 & 0.893 & 0.696 & 11.045 & - & - & - & -     \\
P-attack                          & \underline{-0.061} & \underline{-0.050} & \underline{-0.052} & \underline{53.960} & \underline{48.338}          & \underline{0.194}    & \underline{72.228}     & \underline{0.008} & 0.387   & 0.282   & \underline{0.218} & \underline{39.562} & 18.664          & 0.664 & 43.845 & 	\textbf{0.003}    \\
OUAP  & 0.140   & 0.107   & 0.174  & 46.559   & 25.990           & 0.579  & 70.750        & 0.280   & \underline{0.332} & \underline{0.238} & 0.323   & 36.990   & \underline{27.084}           & \underline{0.560}  & \underline{69.039}    & 0.248   \\
K-attack & 0.842    & 0.842    & 0.650    & 16.962   & 10.097          & 1.015	& 17.250 & 0.081   & 0.588   & 0.568   & 0.420   & 29.374   & 18.828            & 0.723 & 38.616	        & \underline{0.166} \\
UAP  & 0.730    & 0.529    & 0.727    & 20.324   & 11.519            & 1.002 & 27.319 & 	0.351   & 0.620   & {0.444}                          & 0.585   & 24.643   & 13.659            & 0.978 & 37.616    & 0.261   \\
\textbf{SMA (Ours)} & \textbf{-0.716}    & \textbf{-0.570}    & \textbf{-0.748}    & \textbf{70.536}    & \textbf{67.451}           & \textbf{0.007} & \textbf{110.224}      & \textbf{0.004}    & \textbf{-0.494}   & \textbf{-0.308}   & \textbf{-0.648}   & \textbf{62.204}    & \textbf{58.268}    & \textbf{0.083} & \textbf{106.302}       & {\textbf{0.003}}\Bstrut\\ \bottomrule
& \multicolumn{8}{c|}{MANIQA} & \multicolumn{8}{c}{LIQE} \\ 
\cmidrule(lr){2-17}
 & \multicolumn{4}{c|}{Predicted vs. MOS}                  & \multicolumn{3}{c|}{Before vs. After Attack} & \multicolumn{1}{c|}{Invis.}       & \multicolumn{4}{c|}{Predicted vs. MOS} & \multicolumn{3}{c|}{Before vs. After Attack} & \multicolumn{1}{c}{Invis.} \\
& SROCC$\downarrow$                    & KROCC$\downarrow$                    & PLCC$\downarrow$                     & \multicolumn{1}{c|}{RMSE$\uparrow$}                      & Abs.$\uparrow$    & $R\downarrow$        & $\Delta$Rank$\uparrow$ & LPIPS$\downarrow$                  & SROCC$\downarrow$                   & KROCC$\downarrow$                   & PLCC$\downarrow$                    & \multicolumn{1}{c|}{RMSE$\uparrow$}                     & Abs.$\uparrow$    & $R\downarrow$     & $\Delta$Rank$\uparrow$ & LPIPS$\downarrow$\\ \hline %
Clean & 0.827   & 0.849   & 0.633   & 27.425                           & -  & -   & -     & -       & 0.874   & 0.855   & 0.694   & 18.685   & -    & -      & -        & -          \\
P-attack         & \underline{0.141} & \underline{0.098} & \underline{0.046} & \underline{44.559} & \underline{29.769}    & 0.663 & \underline{71.147}   &  \textbf{0.004}    & \underline{0.195} & \underline{0.267} & \underline{0.137} & \underline{45.832} & \underline{36.286} & \underline{0.730} & \underline{64.698} & \textbf{0.004}    \\
OUAP    & 0.182   & 0.128   & 0.236   & 24.717                           & {25.629} & \underline{0.563}  & 68.060 & 0.245   & 0.502   & 0.534   & 0.357   & 28.692   & 25.449   & 0.824   & 47.418 & 0.266   \\
K-attack         & {0.507}                          & {0.491}                          & {0.719}                          & 24.333                           & 7.384    & 1.161 & 26.974                          & \underline{0.166} & 0.698      & 0.707      & 0.521      & 21.014      & 14.061   & 1.112  & 26.944 & \underline{0.166} \\
UAP     & 0.749   & 0.548   & 0.751   & {29.430}                       & 4.308    & 1.384 & 17.198   & 0.261   & 0.800      & 0.776      & 0.608      & 20.838      & 9.576    & 1.202  & 16.961                      & 0.261   \\
\textbf{SMA (Ours)}         & \textbf{-0.439}   & \textbf{-0.244}   & \textbf{-0.418}   & \textbf{44.656} & \textbf{29.788}   & \textbf{0.433} & \textbf{110.336}    & \textbf{0.004}    & \textbf{-0.526}   & \textbf{-0.307}   & \textbf{-0.685}   & \textbf{66.214}   & \textbf{75.025}   & \textbf{0.038}     & \textbf{112.453} & {\textbf{0.004}}\\ \bottomrule
\end{tabular}
}
\end{table*}

Table~\ref{tab:sota_compare} reveals that our method consistently achieves the lowest SROCC among all compared methods while maintaining the largest RMSE. values. For example, when targeting the DBCNN model, P-attack, the best-performing attack among the compared methods, achieves an SROCC of approximately $-0.06$ with an RMSE of nearly $54$. Our method, SMA, decreases the SROCC to under $-0.70$ while increasing the RMSE to over $70$. Notably, all compared methods exhibit their worst attack performance on MANIQA, with the SROCC values of all methods exceeding $0$.
Despite this, SMA still demonstrates superior performance on MANIQA, achieving an SROCC value under $-0.43$,  with an improvement of $0.58$ over the best-performing method in our comparison. 

Interestingly, despite metrics like KROCC and PLCC not being explicitly optimized in the objective function, SMA still exhibits impressive performance with the lowest KROCC, PLCC, $R$, and the largest Abs. values.
On one hand, these results demonstrate that selecting SROCC and MSE as representative metrics for correlation-based and error-based evaluations, respectively, can also optimize other related metrics.
On the other hand, it underscores the effectiveness of the proposed SMA method in attacking both error-based metrics and correlation-based metrics.

Furthermore, from the perspective of disturbing ranking, the largest $\Delta$Rank value of around $110$ is achieved by SMA, which is nearly half of $N$. This suggests that the SMA disturbs the ranking of each image across roughly half of the entire image set on average. It further confirms the effectiveness of SMA in disturbing rankings to a great extent. More importantly, our attack results expose the vulnerability of NR-IQA to maintain ranking consistency, revealing the need for more robust NR-IQA methods in future designs.

\begin{figure}[!t]
    \centering
    \includegraphics[width=.95\linewidth]{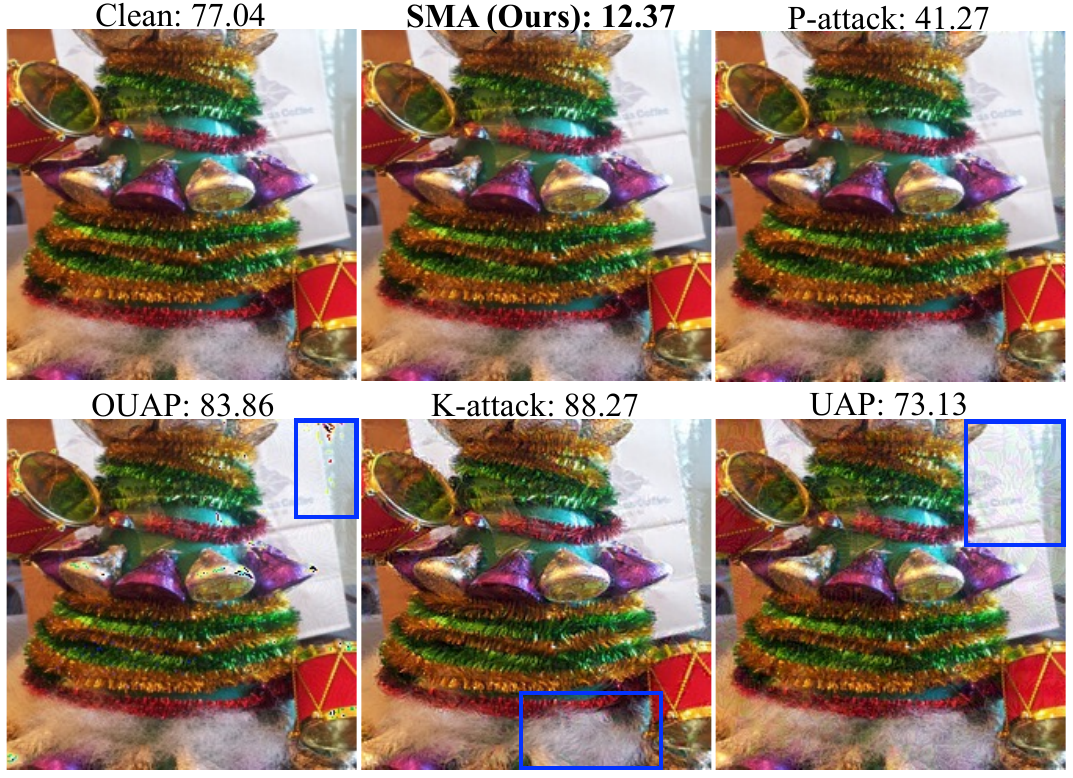}
    \caption{Visualization of adversarial examples on attacking HypyerIQA. The predicted score is on the top of each image. Blue boxes indicate easily noticeable perturbations.}
    \label{fig:visualization}
    \vspace{-6mm}
\end{figure}

For the invisibility of perturbations, Table~\ref{tab:sota_compare} demonstrates that adversarial perturbations generated by our method achieve superior imperceptibility compared to most listed methods.
For a more intuitive comparison, Fig.~\ref{fig:visualization} provides visual results of some adversarial examples. 
From Fig.~\ref{fig:visualization}, perturbations generated by SMA and P-attack are imperceptible, while those introduced by OUAP, K-Attack, and UAP are easily perceivable by human observers (as shown in the blue boxes). 

We also investigate the capability of SMA to manipulate predicted scores to achieve a predefined SROCC value. The experiments show the SROCC between predicted scores of images before and after the attack can approximate the predefined value with a reasonable margin of error.
Such attacks may be crucial in recommendation systems where attackers aim to manipulate the score ordering of adversarial examples to follow a specific order. 
Besides, in addition to SROCC, the optimization of PLCC/KROCC is also explored. The optimization of PLCC generally results in successful attacks, with SMA typically yielding the best outcomes. However, for KROCC, the attack performance is less successful due to the larger differentiable approximation error resulting from the approximation of the sign function within it.

\subsection{Strategy Analysis in Stage One}\label{sec:exper_stage_1}
\begin{figure}[!t]
    \centering
    \includegraphics[width=\linewidth]{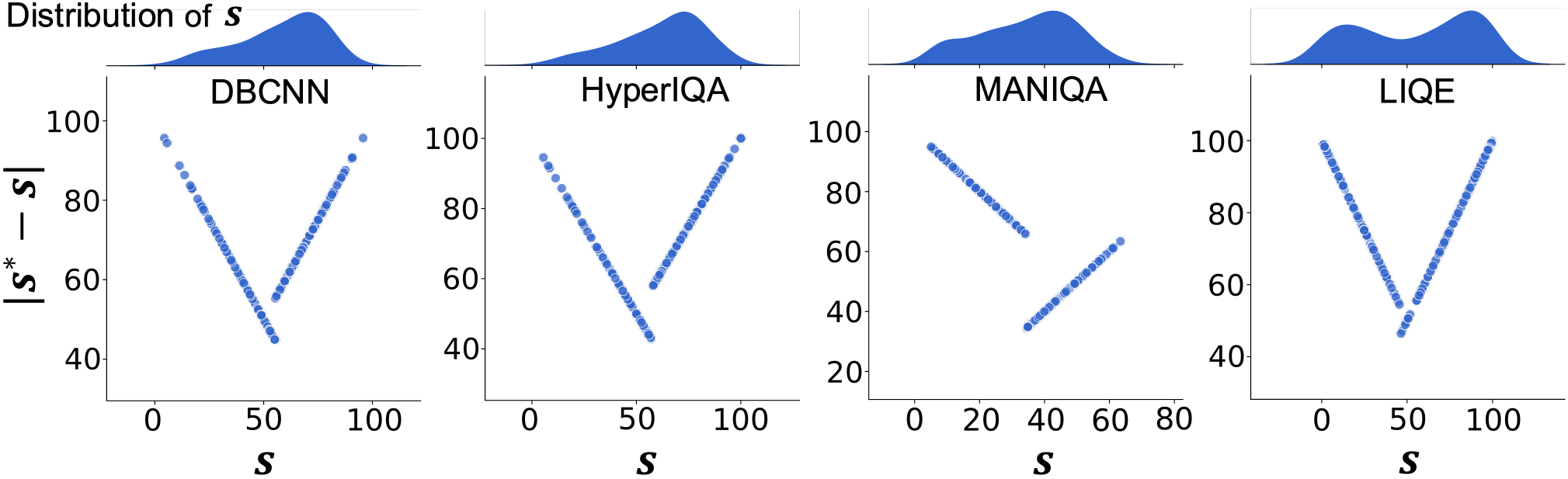}
    \caption{Relationship between original scores $\boldsymbol{s}$ and the score changes $|\boldsymbol{s}^*-\boldsymbol{s}|$ in Stage One.}
    \label{fig:analysis_stage_1}
\end{figure}

In this subsection, we investigate how the optimization strategy in Stage One minimizes the value of SROCC while keeping a large prediction error between $\boldsymbol{s}^*$ and $\boldsymbol{s}$. We calculate the score changes between target scores and original scores, \ie $|s^*_j-s_j|$ for all $j\in\{1,2,\cdots,N\}$, across four NR-IQA models, and present the results in Fig.~\ref{fig:analysis_stage_1}. Meanwhile, the density plot above illustrates the distribution of the original scores $\{s_1,s_2,\cdots,s_N\}$.

From Fig.~\ref{fig:analysis_stage_1}, it is evident that the magnitude of $|\boldsymbol{s}^*-\boldsymbol{s}|$ is closely related to the distribution of $\boldsymbol{s}$. Smaller changes in scores tend to occur around the median of original scores, while larger perturbations are required for scores far from the median of original scores. This observation aligns with intuition, to maximize the change in a single score and disrupt the ranking, a large change is required far from the median to ensure that both the score and its ordering are greatly changed. It implies the optimal magnitude for each sample is related to its position relative to the median in the score set.

\subsection{Error Analysis in Stage Two}
\label{sec:exper_error}

\begin{figure}[!t]
    \centering
    \includegraphics[width=.88\linewidth]{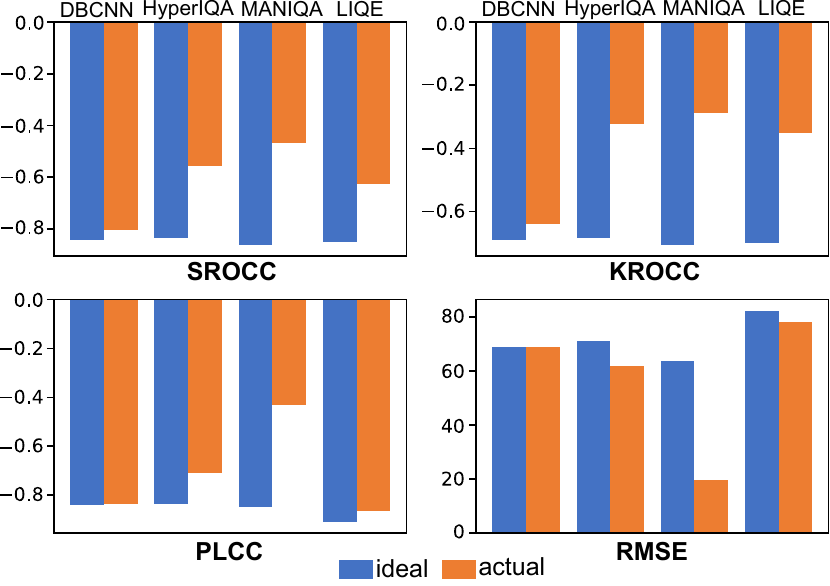}
    \caption{
    The comparison of ideal and actual performance in two stages. Ideal performance is assessed by comparing original scores with target scores obtained in Stage One. Actual performance is evaluated between original scores and predicted scores of adversarial examples generated in Stage Two.}
    \label{fig:error}
    \vspace{-4mm}
\end{figure}

In Stage One of the SMA, we compute a set of target scores with the expectation that the predicted scores of adversarial examples generated in Stage Two will be in proximity to these targets. However, the algorithm employed in Stage Two cannot guarantee exact equality between the predicted scores and target scores. Consequently, it becomes necessary to assess the extent of the gap between the final predicted scores and the desired target scores. To investigate this aspect, we calculate SROCC, KROCC, PLCC, and RMSE between the target scores and original scores. These results serve as the ideal performance achieved in Stage One. Simultaneously, all metrics are computed under original scores and final predicted scores, representing the actual performance obtained in Stage Two. The outcomes are illustrated in Fig.~\ref{fig:error}.

Fig.~\ref{fig:error} reveals that the error between actual performance and ideal performance is relatively small in most cases. Take the DBCNN model as an example, the actual SROCC value is approximately $-0.80$, which is only $0.04$ higher than the ideal value. Similarly, the gap between ideal and actual KROCC/PLCC is less than $0.1$. Remarkably, the actual RMSE closely aligns with the ideal value. The gap of KROCC of the other three NR-IQA models is lower than $0.4$, due to the small changes in scores can easily lead to changes in pairwise comparisons. Nevertheless, the actual KROCC between predicted scores and MOS is still far better than the other attacked methods in Table~\ref{tab:sota_compare}. 
MANIQA is an exception whose actual RMSE is significantly worse than the ideal one. We attribute this phenomenon to the robustness of MANIQA. Even though the target scores for many samples are close to 0, the predicted scores fall within a narrow range of $[5,60]$. This discrepancy results in a significant gap between the target and actual predicted scores. 

\subsection{Ablation Study}
\label{sec:exper_ablation}
There are three main hyperparameters in SMA: $\beta$ in the approximation function $r^\beta(\cdot)$, the multiplier $\lambda_\text{Var}$, and $\lambda_\text{MSE}$ in Eq.~\eqref{eq:objective_stage_1}.
We present ablation studies on them with the target model HyperIQA. 

\noindent \textbf{Approximation function $r^\beta(\cdot)$.}
According to Theorem~\ref{thr:approximation_error}, a small value of $\beta$ is preferable for approximation. However, a small $\beta$ also results to sparse gradients of $r^\beta(\boldsymbol{\tilde{s}})$ with respect to $\boldsymbol{\tilde{s}}$, making it challenging to optimize $\text{SROCC}(\boldsymbol{\tilde{s}},\boldsymbol{s})$ and $\text{MSE}(\boldsymbol{\tilde{s}},\boldsymbol{s})$. Experimental results in Table~\ref{tab:beta} support this point. 
We explore various $\beta$ through $100,000$ optimization steps in Stage One. 
We then present metrics $\text{SROCC}^\beta (\boldsymbol{s}^*,\boldsymbol{s})= 1 - \frac{6} {N(N^2 - 1)}\sum_{j=1}^N \left(r^\beta_j(\boldsymbol{s}^*) - r_j(\boldsymbol{s})\right)^2 $ and $\text{RMSE}(\boldsymbol{s}^*,\boldsymbol{s})$ to assess the optimization efficiency. A smaller $\text{SROCC}^\beta$ and a larger $\text{RMSE}$ indicate faster optimization of SROCC and MSE, respectively.
We also quantify the approximation error as $|\text{SROCC}^\beta(\boldsymbol{s}^{*},\boldsymbol{s})-\text{SROCC}(\boldsymbol{s}^{*},\boldsymbol{s})|$.
From Table~\ref{tab:beta}, as $\beta$ increases, the optimization of the objective in Eq.~\eqref{eq:objective_stage_1} becomes more manageable, resulting in a smaller $\text{SROCC}^\beta(\boldsymbol{s}^*,\boldsymbol{s})$ and a greater $\text{RMSE}(\boldsymbol{s}^*,\boldsymbol{s})$, while the error also generally grows. Nonetheless, the maximum error remains under $0.1$, which is considered acceptable. 

\begin{table}[!t]
\caption{Ablation study of $\beta$ in the approximation function $r^\beta(\cdot)$}
\resizebox{0.75\linewidth}{!}{
\begin{tabular}{lccccc}
\toprule
$\beta$   & 0.2     & 0.4     & 0.6     & 0.8     & 1.0       \\ \hline
$\text{SROCC}^\beta$ & 0.5035 & -0.7691 & -0.8655 & -0.8301 &-0.8316     \\
RMSE    & 27.7607 & 48.6960  & 70.1683 & 70.9954 & 71.0056 \\
error  & 0.0166  & 0.0461  & 0.0605  & 0.0669  & 0.0673 \\
\bottomrule
\end{tabular}
\label{tab:beta}
}
\end{table}

\begin{figure}[!t]  
    \centering
    \includegraphics[width=.9\linewidth]{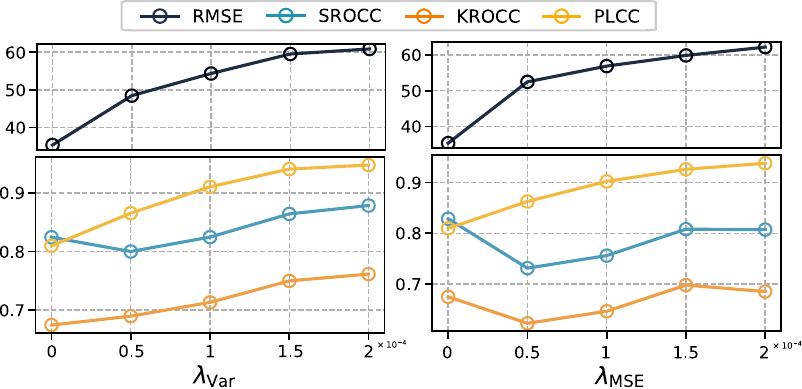}
    \caption{Ablation studies of the multiplier $\lambda_\text{Var}$ and $\lambda_\text{MSE}$.}
    \label{fig:ab_beta_and_lambda}
    \vspace{-8mm}
\end{figure}

\noindent \textbf{The weight $\lambda_\text{Var}$ in Eq.~\eqref{eq:objective_stage_1}.}
A larger $\lambda_\text{Var}$ implies a stricter constraint on the target score variance, making the target easier to achieve in Stage Two. This can be measured by the consistency between the target scores $\boldsymbol{s}^*$ and actual predicted scores $f(\mathcal{I'})$.
Fig.~\ref{fig:ab_beta_and_lambda} (left) confirms this point by illustrating the effect of $\lambda_\text{Var}$ when $\lambda_\text{MSE}=0$. In Fig.~\ref{fig:ab_beta_and_lambda}, the correlations are calculated between $\boldsymbol{s}^*$ and $f(\mathcal{I'})$, and the RMSE is calculated between $\boldsymbol{s}$ and $f(\mathcal{I'})$. With the increase of $\lambda_\text{Var}$, the correlation between $\boldsymbol{s}^*$ and $f(\mathcal{I'})$ generally increased. It confirms the effectiveness of $\text{Var}(\cdot)$ in improving the consistency between target scores and actual predicted scores.

\noindent \textbf{The weight $\lambda_\text{MSE}$ in Eq.~\eqref{eq:objective_stage_1}.} 
A larger $\lambda_\text{MSE}$ implies a more strict constraint on the error between target scores and original scores, thereby enhancing the error between original scores and predicted scores of adversarial examples generated in Stage Two. Fig.~\ref{fig:ab_beta_and_lambda} (right) illustrates the impact of $\lambda_\text{MSE}$ when $\lambda_\text{Var}=0$. As $\lambda_\text{MSE}$ increases, the RMSE between $\boldsymbol{s}$ and $f(\mathcal{I'})$ also increases, confirming the effectiveness of $\text{MSE}(\cdot,\cdot)$ in boosting the error between predicted scores before and after the attack. 
Meanwhile, the increase of $\lambda_\text{MSE}$ will decrease SROCC to some extent.

\section{Conclusion}
This paper introduces a novel framework of correlation-error-based adversarial attacks on NR-IQA models. We present a two-stage method, SMA, which disrupts both individual image scores and intra-set correlations with excellent performance. It highlights NR-IQA's vulnerability in preserving scores and correlations. SMA provides a thorough assessment of robustness in NR-IQA, laying the groundwork for future research on more robust NR-IQA models.

\bibliographystyle{ACM-Reference-Format}
\bibliography{MM}

\end{document}